\documentclass[preprint,12pt,3p]{elsarticle}

\usepackage{amssymb}
\usepackage{float}
\usepackage{amsthm}
\theoremstyle{definition}
\newtheorem{definition}{Definition}[section]
\theoremstyle{remark}

\journal{Journal of Theoretical Biology}

\begin{document}

\begin{frontmatter}

\title{The N-Gene Model for Evolutionary Interaction}

\author[label1,label2]{Tyler Clark\corref{cor1}\fnref{label3}}
\address[label1]{tyler.clark@mail.utoronto.com}


\begin{abstract}
The evolution of altruistic behavior, which confers benefits to others at a cost to the individual, remains a fundamental question in evolutionary biology. While previous models have investigated the conditions favoring the emergence and stability of altruism, they have often employed simplifying assumptions, such as single-gene inheritance and discrete strategies. In this study, we develop a novel evolutionary model that incorporates Mendelian genetics, continuous strategies, and the potential for multiple genes to contribute to a single phenotypic trait. We employ a modified dictator game as the framework for evolutionary interactions and explore the stability of altruistic behavior under various conditions. Our primary result demonstrates that when considering heterozygous genes, altruism can be evolutionarily stable at cost-to-benefit ratios exceeding unity, even with initially low frequencies of altruists in the population. This finding contrasts with the case of homozygous genes, where altruism is only stable at cost-to-benefit ratios greater than 2. We further illustrate that assortative matching, noise in the cost-to-benefit ratio, and the presence of "taking" behavior can influence the evolutionary dynamics of altruism within our model. The generality of our approach allows for its application to a diverse range of evolutionary games and interactions, providing a powerful tool for investigating the emergence and maintenance of social behaviors and personality traits. Our results contribute to the understanding of the evolutionary mechanisms underlying altruism and underscore the importance of incorporating genetic complexity in evolutionary models. This work has implications for the study of social evolution and the genetic architecture of complex behavioral phenotypes.

\end{abstract}


\end{frontmatter}



\section{Introduction}
\label{sec1}

Up until the early twentieth century, Darwin's notion of natural selection was synonymous with 'survival of the fittest': to eat, or be eaten. In the popular imagination at least, this association, or conflation, still persists today. We tend to think of evolution in competitive, selfish terms.

In the later half of the twentieth century, however, evolutionary theories also began to explain non-selfish behaviours by introducing the terms "inclusive fitness" \cite{Ham} and "reciprocal altruism" \cite{reAlt}. In this way, these studies engendered a new way of looking at evolution, one in which survival depended on cooperation instead of precluding it. Work in the late twentieth century also brought into question how much one's genes even impact traits. To what extent are we a \textit{tabula rasa} waiting to be written on? 

With the wide spread use of twin studies, the nature vs. nurture debate has been studied scientifically, and the findings are that, on average, genetics and upbringing are equally significant. This is also roughly the case for altruism and self-interest \cite{altr1}\cite{altr2}\cite{altr3}, although the exact proportions are not necessary for my purposes here; it is simply notable that a significant percentage of the variance is explained via genetics.

One approach to determining the stability of evolutionary strategies is the field of evolutionary game theory. This field relies on the concept of an evolutionarily stable strategy (ESS) \cite{ESS}. An ESS is a modified version of a strict nash equilibrium \cite{NASH}. In brief, this a strategy that cannot be improved after considering the possible strategical space of the other players.

ESSs are motivated entirely differently than those of traditional game theory. It is presumed that these strategies are biologically encoded and heritable. Individuals have no conscious intention in their strategy and are not aware of the game. They reproduce and are subject to the forces of natural selection, with the payoffs of the game representing reproductive biological fitness. Alternative game strategies occasionally occur via a process like mutation. To be an ESS, a strategy must be resistant to these alternatives.

One topic rarely discussed is that there are possible disparities between an ESS within an individual's lifetime and the ESS of a gene over many generations. Often, an analytic solution is found for the ESS of one lifetime, and assumed to be interchangeable with an intergenerational ESS.

This disparity has been noted in parent-offspring conflict \cite{Triv}. In a 1971 paper, for example, Trivers discusses how the optimal parental investment (PI) is different for parents and for offspring. Offspring will want more PI then their siblings, but parents will want to allocate evenly.

The godfather of evolutionary altruism research, W.D. Hamilton, has set the benchmark for research on evolutionary cooperation over the past half a century. He found that \textit{within a single lifetime} the optimal strategy for altruistic giving is based on the inequality $rb>c$ \cite{Ham}, where r is wright's coefficient of relatedness, b is the net fitness benefit to the recipient, and c is the cost to the donor. 

More recently, Alger and Weibull (AW) have investigated this topic at the level of traits and within a modified ESS framework \cite{AW}. However, they again look for an ESS \textit{within a single lifetime}. Both Hamilton's and AW's work only looks at genes that have one allele per loci, also known as homozygeous genes.\\

In this study I consider a model with a class of pairwise interactions that can have gradations of strategy, Mendelian genetics, \footnote{Mendelian genetics refers to meiotic division obeying Mendel's laws} and the possibility of many genes being correlated to a single trait. This allows me to model distributions of traits within a population. I am able to reproduce the results of Hamilton and AW when using homozygeous genes, but when looking at heterozygeous genes I find that the optimal strategy entails a cost-to-benifit ratio of $>1$ for full siblings, counter to their findings.

To explain this intuitively, consider a situation where an individual has a chance to increase their parent’s expected number of offspring rather then maximize their own (Helping their brother at a net loss to themselves, for example). If the cost-to-benefit ratio is $>2$; then both AW and Hamilton would agree that this is not a stable strategy. However, from the parent’s point of view, it is optimal for their offspring to help one another at a cost-to-benefit $\alpha >1$.  When an individual has offspring, the initially negative altruism cost pays off through the generations as they enjoy a higher expected number of children then their short-term selfish counterparts.

\subsection{Model Introduction}
Consider a population of sexually reproducing individuals who are matched into pairs. Each pair is engaged in some interaction. Each individual carries some heritable trait $\theta \in T$, where T is a set of potential traits. Here we use the word 'trait' in a general sense, allowing it to mean anything that may influence the outcome of the interaction, such as intelligence, size, decision rule, etc.

This is the game, and after it is played the individuals survive until replication with a certain probability which is impacted during the game.

\subsubsection{Fitness}
Many models deal with the fitness associated with a given trait in an abstract sense, as being the expected number of offspring an individual creates. If we assume that an individual's life can be broken into two parts, immaturity and maturity, once the individual lives until maturity, they will give birth without possibility of death.

When looked at in a probabilistic frame, we see that changes in genotype fitness without evolutionary forces/interactions can either act on the number of offspring ($n$) an individual has or the chances the individual has to live until reproduction/maturity ($p$). \\

\theoremstyle{definition}
\begin{definition}{\textit{Mature Fitness}}
is the number of offspring (n) an individual about to reproduce has that live until reproduction without evolutionary forces, where each $i^{th}$ offspring has a probability ($p_i$) of living until this point.

\begin{equation}
    \omega=\sum_{i=1}^n p_i
\end{equation}
\end{definition}

\theoremstyle{definition}
\begin{definition}{\textit{Immature Fitness}}
is the expected number of offspring an individual that lives until reproductive age has, taking into account that the individual has not yet reached reproductive age, and will do so with a probability (p).

\begin{equation}
    \omega'=p\omega
\end{equation}
\end{definition}

The above formula for mature fitness is the expectation of the binomial poisson distribution, which in the limiting case that all offspring have an equal probability of living until replication reduces to $\omega=np$.

\theoremstyle{definition}
\begin{definition}{\textit{Evolutionary Interaction}}
is the immature fitness resulting from the interaction of an individual with trait ($\theta$) with another individual with trait ($\theta'$) .

\begin{equation}
    I(\theta,\theta')=(p+\delta p)(\omega+\delta \omega)
\end{equation}
\end{definition}

Because trait inheritance is a distribution, not deterministic, in this model it is easiest  to specify a constant mature fitness $\omega_c$ which all individuals in the gene pool possess.\\

We can define the change in probability of living until replication after the interaction as

\begin{equation}
    f(\theta,\theta')=p+\delta p
\end{equation}

\begin{figure}[H]
\caption{This figure shows how two first generation parents undergo meiotic reproduction in order to produce two offspring. The offspring then play an evolutionary game which changes their chance of living until reproduction. The second generation then undergoes meiotic reproduction to create the third generation. }
\centering
\includegraphics[width=0.9\textwidth]{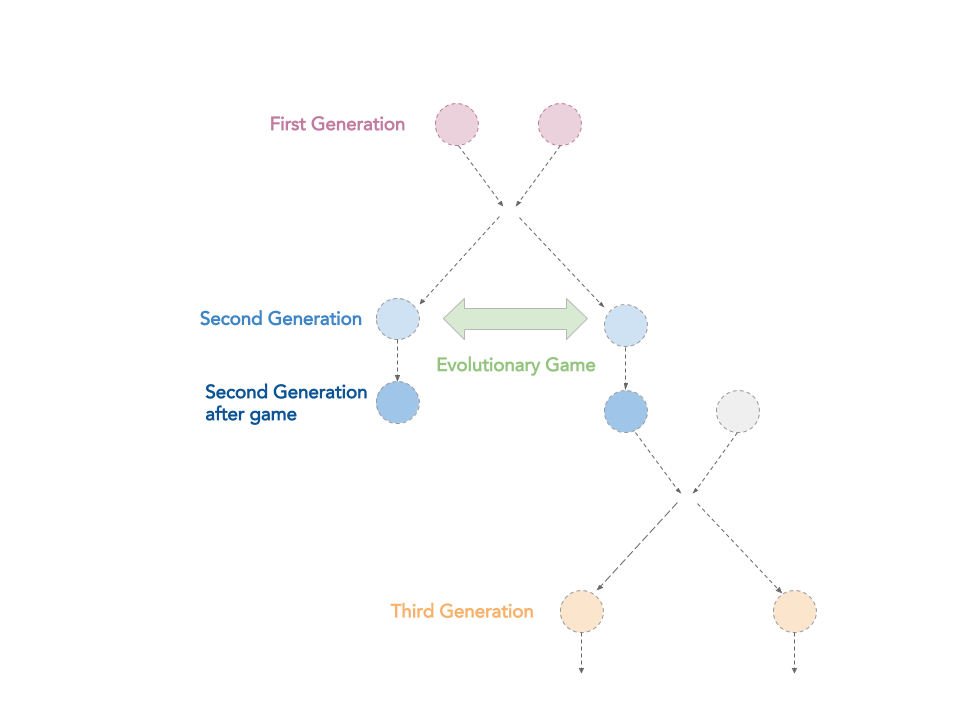}
\end{figure}

\subsubsection{Traits as Macrostates}

Let each individual have a set of n genes that are correlated to the trait $\theta$. Let  the trait property be a function of the individual's total number of correlated genes, which we will call the genomic macrostate. This can be contrasted with the individual's specific set of genes, or the genomic microstate.\\

\theoremstyle{definition}
\begin{definition}{\textit{Trait} $\theta(k)$}
is the phenotypic expression of the genomic macrostate $k$ of the individual; which effects the strategy of an evolutionary interaction.
\end{definition}

\subsubsection{Modified Dictator Game}
Let us consider an asymmetric two player game that is a modified version of the dictator game. Let Player 1 be the dictator who has the ability to give Player 2 a part of their total fitness with a benefit-to-cost ratio of $\alpha$. The strategy set for the game would be \\

$(p,p) \mapsto (f(\theta',\theta),f(\theta,\theta'))=(p-\delta p,p+\delta p \alpha )$ \\

Where $\delta p \alpha \leq (1-p)$ and $\delta p<p$

We can define the mapping of the genomic macrostate to the trait as

\begin{equation}
    \theta(\Omega)=a\Omega+b
\end{equation}

Where $a$ specifies the magnitude of impact the interaction has on fitness and $b$ changes the minimum amount of giving required.

\subsection{Assumptions of the Model}

\begin{enumerate}
    \item Let there be $n$ genes 
    \item Let the gene-pool for each gene have two competing alleles. One allele has a positive influence on the trait, and the other is normalized to have no effect.
    \item Let each gene have two alleles, giving rise to the possibility of heterozygosity.
    \item Let the frequency of the correlated allele in each of the $n$ genes be equal.
    \item Mendelian inheritance applies.
    \item Let the correlated gene be dominant with a probability $p$.
    \item Let each gene be equally correlated with the trait.
\end{enumerate}

In this application let the genomic microstate of an individual $s= [g_1,g_2...g_n]$ be the specific set of alleles that an individual possesses. Due to the fact that all genes have an identical effect on the trait, the relevant factor is not which genes have the correlated allele, but the total number of correlated alleles.\\

Let $\theta(\Omega)$ be the trait as a function of the individual's macrostate $\Omega=\sum^n_{i=1}g_i$ where $g_i=(1,0)$ is a set indicating whether the gene is correlated to the trait or not.
If $g_i=(1,1)$ the trait is correlated, if $g_i=(0,0)$ uncorrelated, and if  $g_i=(1,0)$ then the trait is correlated with a probability $p$, which is the chance of being dominant. \\

Instead of calculating the macrostate directly, one can also do this with knowledge of the frequencies of the homozygeous correlated alleles $\rho_{BB}$, the heterzygeous set of alleles $\rho_{BW}$, and the homozygeous uncorrelated set of alleles $\rho_{WW}$. Because the genes are identical, the previous frequencies are equal for each gene.

\subsection{Mating}

\begin{figure}[H]
\centering
\includegraphics[width=0.9\textwidth]{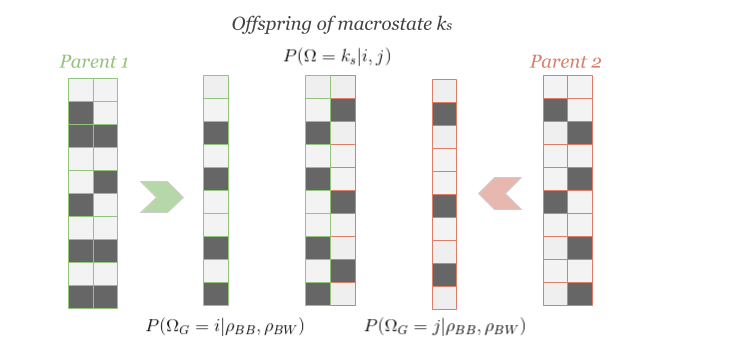}
\end{figure}

Figure 1 shows the intuition behind the process of mating in this model.
Given parents of macrostate $i,j$, and assuming Mendel's law of segregation holds, we can calculate the macrostate of the gamete $\Omega_G$ that contains only a single allele per gene.

Using the equations shown below, one can derive probability or frequency estimate of how many individuals will have a certain macrostate, and therefor how many parents of macrostates $i,j$ will mate together and the probability of their child having macrostate $k_s$

\subsubsection{Heterozygeous populations}

 We can create a tensor
\begin{equation}
    \Gamma=\sum_{i,j,k...}P(\Omega_G=i|\rho_{BB},\rho_{BW})P(\Omega_G=j|\rho_{BB},\rho_{BW})\prod_{s=1}^N P(\Omega=k_s|i,j)
\end{equation}

Which specifies the frequency of N offspring  with ordered macrostates $k_{1,2..N}$, given parents with gamete macrostates i, \ j.\\

First consider the last term, $ P(\Omega_G=k_s|i,j)$. This is the probability of having an offspring with macrostate $k_s$ given parents of gamete macrostates \textit{i}, \ \textit{j.} \\

We find that when the correlated gene is completely recessive:
\begin{equation}\label{eq:1}
    P(\Omega=k_s|i,j)_{recess}= {n-j \choose i-k_s}{j \choose k_s}\frac{k_s!(n-i)!}{n!} \ \  \forall i>j
\end{equation}

Where n is the number of correlated genes. We can see that if $j>i$ the equation is the same with \textit{j} and\textit{ i} exchanged. 

\subsubsection{Dominance}
When the correlated gene is dominant with a certain probability $p$ we arrive at the following equation:

\begin{equation}
    P(\Omega=k_s|i,j,p)=\sum_{v=0}^{k_s}P(\Omega=k_s-v|i,j)_{recess}{j+i-2k_s+2v \choose v}p^v(1-p)^{i+j-2k_s+2v}
\end{equation}

$\forall$\\
a) \ $k_s-v \leq j$ \\
b) \ $ i+j-2k_s+2v \geq v$\\
c) \ $n \geq i+j-2k_s+2v$\\
d) \ $k_s-v\geq i+j-n$\\

If each individual has\textit{ n} correlated genes; and each gene loci has the possibility of being either heterozygeous or homozygous\footnote{In the case that the individual's gene is homozygeous, it can either be correlated or uncorrelated}, the expected gamete macrostate from choosing one allele from each gene in which to reproduce can be expressed using the binomial formula below.

\begin{equation}\label{eq:makedist}
    Pr(\Omega_G=k|\rho_{BB},\rho_{BW})={n\choose k}(\rho_{BB}+\frac{\rho_{BW}}{2})^{k}(1-\rho_{BB}-\frac{\rho_{BW}}{2})^{n-k}
\end{equation}

As you can see, the exact macrostate of each parent does not have to be known: just the macrostate probability of the single-allele gamete that is exchanged during mating.

\subsubsection{Homozygeous and Heterozygeous Frequencies}
After inspecting $P(\Omega=k_s|i,j)_{recess}$ we can see this implies that there are $H_{cor}=k_s$ Homozygeous correlated genes, $H_t=i+j-2k$ Heterozygeous genes, and $H_{un}=n-i-j+k$ Homozygeous uncorrelated genes.

Therefor, the expected frequency of correlated or uncorrelated genes going into the next generation for two offspring \textit{without evolutionary pressures} is:

\begin{equation}
    E[H]=\sum_{i,j,k_1,k_2}P(\Omega_G=i|\rho_{BB},\rho_{BW})P(\Omega_G=j|\rho_{BB},\rho_{BW})P(\Omega=k_1|i,j,p)P'(\Omega=k_2|i,j,p)
\end{equation}

 Where
 
 \begin{equation}
    P'(\Omega=k_s|i,j,p)=\sum_{v=0}^{k_s}P(\Omega=k_s-v|i,j)_{recess}{j+i-2k_s+2v \choose v}p^v(1-p)^{i+j-2k_s+2v}H(n,i,j,k_s)
 \end{equation}

 The above notation can be simplified to
 
\begin{equation}\label{eq:gammaH}
    E[H]=\sum_{i,j,k_1,k_2} \Psi(i,j,k_1,k_2,p)
\end{equation}

We would expect this result to replicate the Hardy-Weinburg equilibrium which, using simulations, we find it does. 

\subsection{The Pairwise Interaction}

Due to the symmetry between both offspring in equation  \ref{eq:gammaH} we can define 

\begin{equation}\label{eq:HcondF}
    E[H|F]=\sum_{i,j,k_1,k_2} \Psi(i,j,k_1,k_2,p)F(\theta(k_1),\theta(k_2))
\end{equation} 

Where

\begin{equation}
   F(\theta(k_1),\theta(k_2))=f(\theta(k_1),\theta(k_1))+f(\theta(k_1),\theta(k_2))
\end{equation}

\subsection{Modeling}

We can use equation \ref{eq:HcondF} to determine the frequencies of the correlated and uncorrelated genes of the $t+1$ season. Then we can re-derive the distribution using equation \ref{eq:makedist}.

\section{Results}
Using simulations, the following important results were found:

\subsection{Main Results}

\begin{enumerate}

    \item Homozygeous genes (single allele per loci) found altruism to be evolutionarily stable for cost-to-benefit ratios $>2$.
    \item Heterozygeous genes (two alleles per loci) found altruism to be evolutionarily stable for cost-to-benefit ratios of $>1$ with very low initial populations of altruists.
    
    \item When considering portions of random assortative matching, the cost-to-benefit ratio was equal to the inverse of the probability of being matched with the random individual in the population compared to a sibling.
\end{enumerate}

\subsection{Secondary Results}
\begin{enumerate}
    \item Hardy-Weinberg Equilibrium was found for no evolutionary pressures with heterozygeous genes, but \textit{not} for homozygeous genes. Initial conditions only effected the mean of the distribution.
    
    \item Altruism was also stable when taking was allowed, although the process was slower and the dynamics changed.
    
    \item When mean centered stochastic noise is added to a cost-to-benefit ratio equal to 1, then altruism is preferred.
\end{enumerate}

\section{Discussion}

The present study extends previous work on the evolution of altruistic behavior by introducing a novel model that incorporates Mendelian genetics, gradations of strategy, and the potential for multiple genes to contribute to a single phenotypic trait. The most significant finding is that when considering heterozygous genes, altruistic behavior can be evolutionarily stable at cost-to-benefit ratios exceeding unity, even with initially low frequencies of altruists in the population. This result stands in contrast to the findings for homozygous genes, which only support the stability of altruism at cost-to-benefit ratios greater than 2.

The approach employed in this study, which models traits as macrostates determined by the frequencies of correlated alleles across multiple genes, provides a useful framework for capturing the distribution of traits within a population. The linear mapping from genomic macrostate to expressed trait serves as a reasonable first approximation.

The use of a modified dictator game as the evolutionary interaction effectively models altruistic versus selfish behavior with adjustable costs and benefits. However, the model's flexibility allows for its application to a wide range of evolutionary games and interactions. For example, the model could be adapted to study the evolution of cooperation in the prisoner's dilemma, the emergence of fairness norms in the ultimatum game, or the dynamics of reciprocity in the repeated prisoner's dilemma. By adjusting the payoff structure and the mapping from macrostates to traits, researchers can investigate the evolutionary stability of various strategies in different contexts.

Moreover, the model's ability to incorporate multiple genes and gradations of traits makes it well-suited for studying the evolution of complex social behaviors and personality traits. For instance, the model could be used to explore the evolutionary basis of empathy, trust, or aggression by considering the interplay of multiple genetic factors and environmental influences. Such investigations could shed light on the origins and maintenance of individual differences in social behavior and contribute to our understanding of the genetic architecture underlying human psychology.

The derivations of offspring macrostates arising from the macrostates of parental gametes, considering both recessive and dominant inheritance patterns, appear rigorous. The model's ability to replicate the Hardy-Weinberg equilibrium in the absence of evolutionary pressures serves as a valuable validation.

Additional notable results include the finding that assortative matching enables the stability of higher cost-to-benefit ratios in proportion to the ratio of matching with siblings versus the general population, the confirmation that altruism can still evolve even when "taking" behavior is allowed, albeit at a slower pace, and the demonstration that mean-centered noise added to the cost-to-benefit ratio favors altruism.

Several avenues for further discussion and extension of this work present themselves. First, investigating the sensitivity of the results to the assumed linear mapping between macrostate and trait, and exploring the consequences of nonlinear mappings, could yield valuable insights. Second, relaxing the assumption that all genes are equally correlated with the trait may provide a more nuanced understanding of the system. Third, considering more complex interactions or multi-round games could reveal additional factors influencing the evolution of altruism. Finally, conducting larger simulations with more genes, finer gradations of traits, and larger population sizes could help validate the scalability of the results.

In conclusion, this study represents a valuable contribution to our understanding of the evolutionary mechanisms underlying the emergence of altruistic behaviors. The modeling of traits as genomic macrostates with Mendelian inheritance provides a useful framework, and the key result that heterozygosity allows altruism to evolve under a wider range of conditions is both intriguing and warrants further investigation. The model's flexibility and potential for application to various games and interactions make it a promising tool for future research on the evolution of social behavior and personality traits. With additional expansion and more detailed analysis, this work has the potential to make a significant impact in the field.

\begin{figure}[H]
\caption{This figure shows how the distribution of traits in a population changes depending on pressures from the evolutionary game. The x-axis represents the genomic macrostate, the y-axis represents the season of the population. The 0 macrostate represents no giving in the dictator game, whereas 3 represents giving to the highest possible amount. Specifically, the cost-to-benefit ratio $\alpha$ was 4, and the initial frequency of genes was $97.5 \% $ selfish.  }
\includegraphics[width=0.8\textwidth]{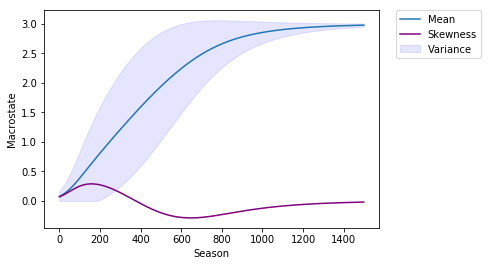}
\end{figure}

\begin{figure}[H]
\caption{This figure shows how the distribution of traits in a population changes during pressures from the evolutionary game. The x-axis represents the genomic macrostate frequency, the y-axis represents the season of the population. The 0 macrostate represents no giving in the dictator game; whereas 3 represents giving to the highest possible amount. Specifically the cost to benefit ratio $\alpha$ was 4, and the initial frequency of genes was $97.5 \% $ selfish  }
\includegraphics[width=0.8\textwidth]{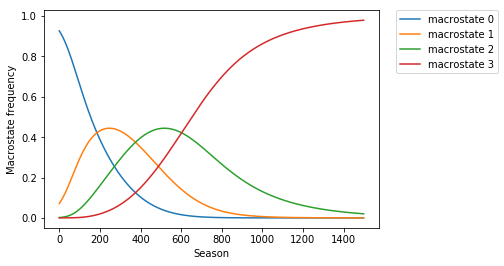}
\end{figure}






\bibliographystyle{elsarticle-num}

\bibliography{sample}

\end{document}